\documentclass{article}

\PassOptionsToPackage{numbers, compress}{natbib}
\usepackage[preprint,creativeai]{neurips_2026}

\usepackage[utf8]{inputenc}
\usepackage[T1]{fontenc}
\usepackage{hyperref}
\usepackage{url}
\usepackage{booktabs}
\usepackage{amsfonts}
\usepackage{nicefrac}
\usepackage{microtype}
\usepackage{xcolor}
\usepackage{graphicx}

\title{Smell with Genji: Rediscovering Sensory Experience with an AI Co-Smelling Partner}

\author{%
  Vera Yu Wu\thanks{These authors contributed equally to this research.} \\
  MIT Media Lab\\
  \texttt{verawu@mit.edu} \\
  \And
  Awu Chen\footnotemark[1] \\
  MIT Media Lab\\
  \texttt{awwu@mit.edu} \\
  \And
  Yunge Wen \\
  New York University\\
  \texttt{yw3776@nyu.edu} \\
  \And
  Yaluo Wang \\
  Harvard University\\
  \texttt{yaluo\_wang@mde.harvard.edu} \\
  \And
  Jiaxuan Olivia Yin \\
  Independent Researcher\\
  \texttt{oliviayinlab@gmail.com} \\
  \And
  Yichen Wang \\
  Harvard University\\
  \texttt{yichen\_wang@gsd.harvard.edu} \\
  \And
  Qian Xiang \\
  Harvard University\\
  \texttt{qian\_xiang@gsd.harvard.edu} \\
  \And
  Haoxi Zhang \\
  MIT Media Lab\\
  \texttt{haoxi@mit.edu} \\
  \And
  Paul Pu Liang \\
  MIT Media Lab\\
  \texttt{ppliang@mit.edu} \\
  \And
  Hiroshi Ishii \\
  MIT Media Lab\\
  \texttt{ishii@mit.edu} \\
}

\usepackage{float}
\begin{document}

\maketitle

\begin{figure}[!htbp]
\centering
\includegraphics[width=\linewidth]{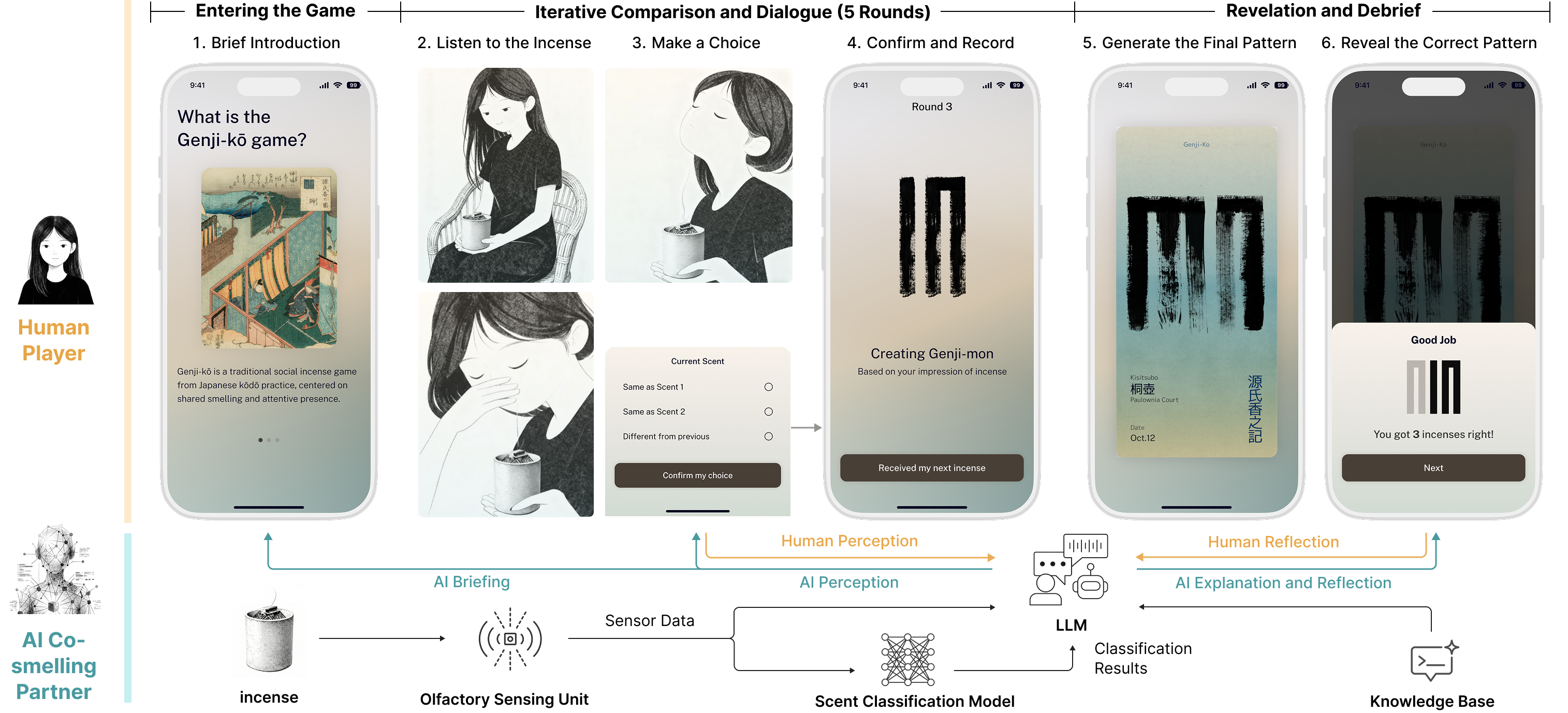}
\caption{\textbf{Overview of \emph{Smell with Genji}.} (Top) Mobile interface for guided gameplay and real-time \emph{Genji-mon} updates. (Bottom) AI co-smelling partner pipeline translating olfactory sensing data and classification results into reflective LLM dialogue.}
\label{fig:probe}
\end{figure}

\begin{abstract}
Olfaction contributes substantially to human perception, yet the olfactory-verbal gap often leaves people struggling to articulate what they smell. We present \emph{Smell with Genji}, an AI-mediated interactive experience adapting \emph{Genji-k\=o}, a traditional Japanese scent-matching game. This practice maps comparative judgments across a sequence of scents onto \emph{Genji-mon}, providing visual patterns as shared reference points for communication. Within this framework, we introduce an AI co-smelling partner integrating multichannel olfactory sensing, a Transformer-based classification model, and a large language model (LLM)-driven conversational interface to translate sensor signals into descriptive dialogue. Rather than deploying machine olfaction purely for identification, the system positions the AI as a co-present partner sharing its sensory impressions while human judgment remains open. Placing the machine's perspective alongside human intuition creates a space for shared reflection, encouraging participants to deliberately evaluate and articulate their choices. This work explores how computational systems can support sensory interaction grounded in dialogue, fostering reflective articulation and preserving human agency in making sense of the olfactory world.
\end{abstract}

\section{Introduction}

Olfactory perception is closely linked to affective and memory-related processes and often bypasses high-level cognitive filtering \cite{danthiir2001, herz1996}. Cultivating an awareness of smell has been shown to enrich daily experiences of mood and memory, offering tangible benefits for personal well-being \cite{bratman2024, licon2018}. However, this potential remains difficult to leverage due to a pervasive olfactory-verbal gap \cite{stevenson2007}. Compared to other sensory modalities, references to smell are infrequent across languages and communicative contexts, reflecting a limited shared vocabulary for describing olfactory experiences \cite{majid2014, olofsson2015}. As a result, while people readily acknowledge the presence of an odor, they often struggle to articulate what they are smelling \cite{keller2004}.

Cultural practices centered on fragrance offer a viable framework to address this limitation. \emph{Genji-k\=o}, a traditional social olfactory game rooted in Japanese \emph{K\=od\=o}, shifts the interaction from direct verbal description toward relative, comparative discrimination. This structure renders subjective sensory impressions more accessible for comparison and open to further empirical exploration.

Recent developments in AI open up new possibilities for computational systems to participate in such sensory practices. While LLMs excel at supporting interpretation and dialogue \cite{shaer2024, yang2024}, they lack direct olfactory perception. Conversely, advances in machine olfaction demonstrate that scents can be computationally encoded from physical sensor signals \cite{feng2025}. Bridging these modalities allows a system to ground classification results and sensor-derived representations in language, enabling meaningful multi-sensory dialogue with human participants.

We present \emph{Smell with Genji}, an olfactory experience in which an AI co-smelling partner takes the seat of a fellow perceiver in the traditional \emph{Genji-k\=o} practice (Figure \ref{fig:probe}). Subtle scent discrimination is hard for people and for gas sensing models alike \cite{cain1979, furizal2023}, so the AI acts not as an authority but as a second perceiver, making each session an act of joint sensemaking. Points of alignment and divergence between the two readings push vague impressions toward deliberate articulation. The system thereby expands human agency in making sense of the olfactory world and invites reflection on perception itself.

\section{Related Work}

\subsection{Olfactory Interaction and \emph{Genji-k\=o}}

Olfaction remains comparatively underexplored in interaction research, which concentrates on visual and auditory modalities \cite{brooks2023, obrist2016}, and where scent does appear it is usually an output. Olfactory displays deliver fragrance to influence mood or cognitive performance \cite{amores2017}, and many systems pair a given input with a scent prepared in advance \cite{holloman2022}. More recent work has begun to incorporate language models. OdorAgent pairs an LLM with a text-to-image model to select and schedule odors for film \cite{zhang2024odor}. The Anemoia Device extends this to composition, where a dial-based interface guides the model from an archival photograph to a narrative and then to a scent formula \cite{clarke2025}. In each case the scent is what the computation delivers.

A different relation to scent appears in \emph{Genji-k\=o}, a cultural practice in which the scent is present from the outset and several perceivers each make sense of it. Participants sequentially ``listen to'' five incense samples, judge each against the ones before it, and record the result as a \emph{Genji-mon}, five vertical lines with those judged identical joined at the top \cite{looney2024}. Samples may repeat, so the answer is a grouping of the five, not a naming of each, and fifty-two groupings are possible. The notation turns a private impression into a shared reference point, making visible where perceivers align and where they diverge. Prior work digitized these mechanics with a computer-controlled censer and a tablet interface for solitary play \cite{yokokubo2019}, which isolated the matching task from its core social exchange.

\subsection{Machine Olfaction}

Machine olfaction asks what a machine can perceive from a scent that is already present. Electronic nose systems discriminate volatile profiles using gas sensor arrays, though they face persistent difficulties with sensor drift and chemically similar samples \cite{furizal2023}. Extending this line beyond controlled laboratory conditions, a large-scale dataset of everyday substances reports a marked drop in accuracy when recognition moves from recorded signals to live sampling \cite{feng2025}. A parallel direction approaches scent through language, asking whether language models carry usable olfactory structure learned from text alone \cite{kurfali2025}. Interactive work has begun to probe the same question, having participants sniff and describe a scent for an LLM to identify, with alignment reported as limited and uneven across scents \cite{zhong_sniff_2024}. Across both approaches, the machine's reading is judged against a reference standard, whether a chemical label or a human rating. Placed inside an interaction rather than an evaluation, our system occupies a different, more conversational position.

\subsection{From Reliance to Reflection}

Research on AI-assisted decision-making treats appropriate reliance, between overreliance and underreliance, as the design objective \cite{schemmer2023}, and finds that forcing deliberation before a decision reduces uncritical acceptance \cite{bucinca2021}. How an output is positioned therefore matters independently of its accuracy. Creative AI research makes the same case aesthetically, treating the interpretive divergence introduced by a language-mediated pipeline as a way to see a remembered movement afresh \cite{liu2025}. In our system that deliberation is applied to a perceptual judgment rather than a decision. Participant and machine smell the same incense, and the participant commits to a comparative judgment before hearing the AI's. Neither perceiver has privileged access to the scent, so the AI's reading prompts the participant to revisit that judgment and articulate it to a partner that answers.

\section{System Architecture}

\begin{figure}[H]
\centering
\includegraphics[width=0.80\linewidth]{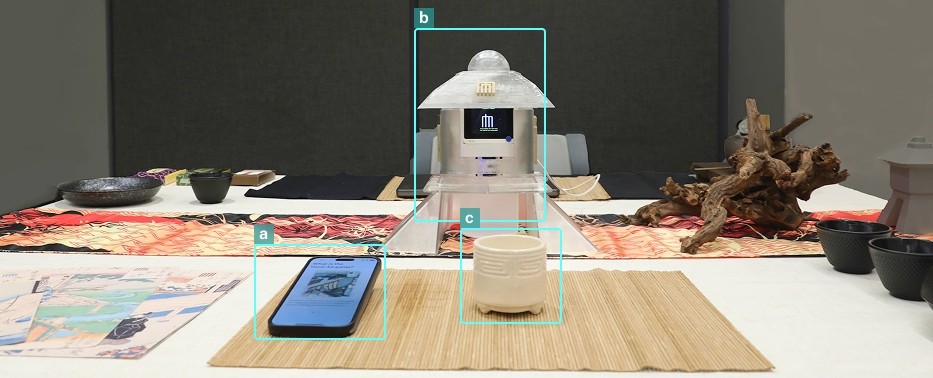}
\caption{The olfactory game setup. (a) Mobile application for guided gameplay and \emph{Genji-mon} visualization. (b) AI co-smelling partner embodied in a 3D-printed enclosure with sensing and display. (c) Incense holder for smelling.}
\label{fig:setup}
\end{figure}

The \emph{Smell with Genji} architecture integrates an olfactory game setup and an AI co-smelling partner embodied in a custom enclosure (Figure \ref{fig:setup}). Physical sensing, interactive visualization, and LLM-driven reasoning operate synchronously across a distributed workflow.

\subsection{Olfactory Game Setup}

The interaction takes place on a low table inspired by traditional \emph{K\=od\=o} practice, co-locating the incense holder, mobile interface, and AI partner on a shared surface so that participant and machine draw on the same incense. Positioned centrally, the AI partner's physical enclosure provides a spatial presence analogous to a human co-player. The setup also accommodates several participants in one session, each keeping an independent record and comparing readings with one another and with the AI partner. The mobile application, developed in React with WebSockets, serves as the control layer. At session initialization, scanning a QR code pairs the physical incense sequence with the digital system, aligning the AI partner with the corresponding scent data.

\subsection{AI Co-Smelling Partner}

\begin{figure}[!htbp]
\centering
\includegraphics[width=\linewidth]{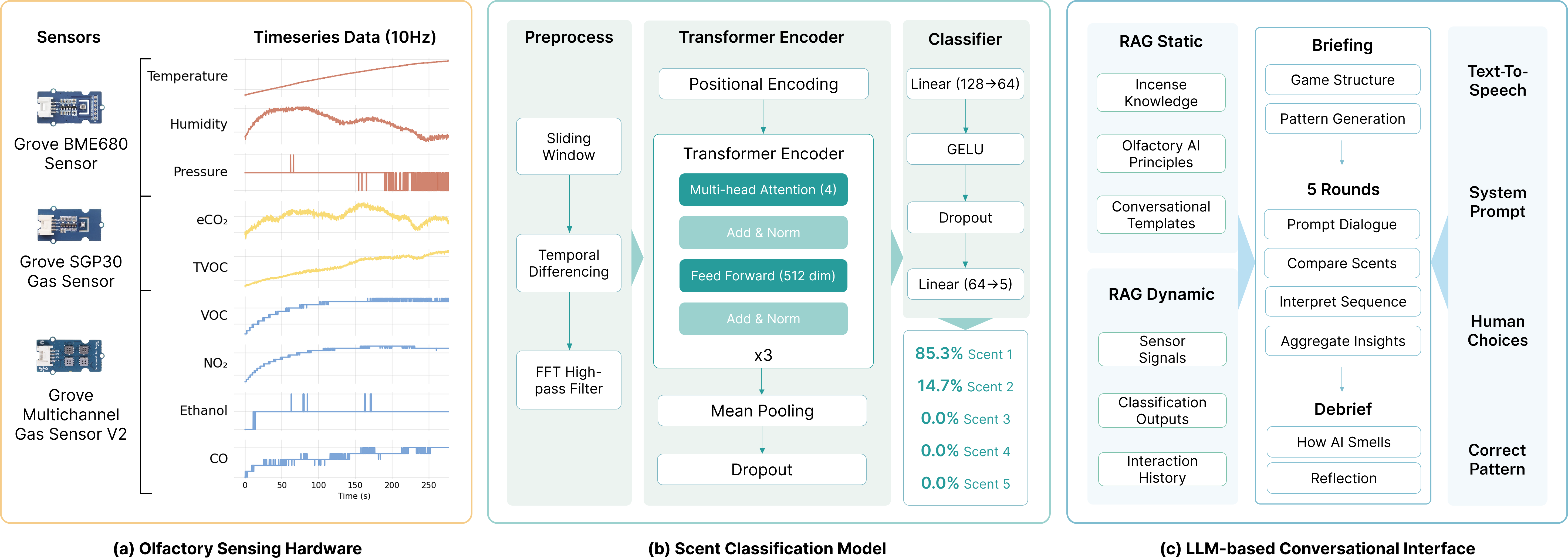}
\caption{The AI co-smelling partner. (a) Olfactory sensing hardware: three gas sensors collect nine channels of time-series signals. (b) Scent classification model: a Transformer encoder processes the sensing data to classify scents. (c) LLM-based conversational interface: a RAG pipeline ingests classification results and sensing data to synthesize dialogue across the briefing, five game rounds, and debrief.}
\label{fig:partner}
\end{figure}

The AI partner operates as a sensing, reasoning, and dialogue pipeline (Figure \ref{fig:partner}).

\paragraph{Olfactory Sensing Hardware.}
The sensing module captures nine channels of environmental and gas composition data via three sensors: a Grove BME680 (temperature, humidity, pressure), a Grove SGP30 (TVOC, eCO$_2$), and a Multichannel Gas Sensor V2 (VOC, NO$_2$, ethanol, CO). To establish baseline data without combustion, 10 g of incense material per sample was sealed with sensors positioned 20 mm away, preventing smoke contamination and preserving sensor longevity. Sampling proceeded at 10 Hz over 5-minute windows across indoor/outdoor and morning/evening conditions, accumulating 75 minutes of recording and 405,000 data points across 9 signal channels.

\paragraph{Scent Classification Model.}
Feature extraction was evaluated using temporal differencing, high-pass FFT filtering, standard scaling, and five window-stride configurations. The underlying model couples a Transformer encoder backbone with an MLP classifier head for 5-class incense categorization, processing the raw gas sensing signals. Under controlled conditions, the model achieves approximately 40\% accuracy against a 20\% chance baseline, with errors arising from overlapping volatile organic profiles among chemically similar samples. Real-time inference executes via sliding windows with accumulative voting. The Transformer encoder processes signal trajectories over time instead of static snapshots, computationally capturing the dynamic temporal unfolding of traditional \emph{K\=od\=o} evaluation.

\paragraph{LLM-based Conversational Interface.}
The dialogic layer bridges machine sensing with natural language using GPT-4o via a dual Retrieval-Augmented Generation (RAG) framework. A static vector database stores domain knowledge, olfactory AI principles, and dialogue templates; a dynamic memory module tracks gas sensing data, classification results, and cross-session interaction history. Guided by a system prompt, the agent ingests the model's classification results and underlying signal trends, translating them into contextually grounded descriptions with mode-specific retrieval adapting prompts across the briefing, five game rounds, and debriefing phases. A custom persona and text-to-speech (TTS) module deliver a consistent voice partner. The prompt constrains the agent to frame its signal derivations as subjective interpretations, not objective facts, while requiring participants to commit to their comparative judgment before the AI reveals its output.

\section{Interactive Experience}

\begin{figure}[!htbp]
\centering
\includegraphics[width=\linewidth]{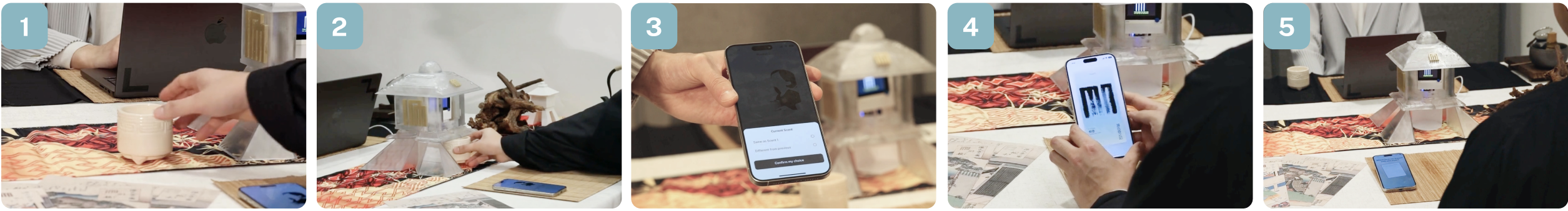}
\caption{The interactive flow of \emph{Smell with Genji}. The sequence loops through (1) ``listening'' to the incense, (2) engaging in AI-guided dialogue, (3) confirming comparative judgments, and (4) rendering the evolving \emph{Genji-mon} patterns, concluding with (5) an AI debriefing.}
\label{fig:process}
\end{figure}

The interaction is structured as a five-round ``incense listening'' sequence (Figure \ref{fig:process}).

\paragraph{Entering the Game.}
Participants first learn the structure of \emph{Genji-k\=o} through the mobile application and a spoken AI briefing. A brief calibration sequence then presents the five incense samples in turn to establish an olfactory baseline, encouraging participants to attend to each scent before formal gameplay starts.

\paragraph{Iterative Comparison and Dialogue.}
The core interaction consists of five sequential rounds. The first round establishes a baseline scent. From the second round onward, the participant first ``listens to'' the current incense and indicates on the mobile application whether the scent matches an earlier sample or represents a distinct fragrance. 

Once this initial choice is made, the participant places the same incense holder they have just smelled from into the base of the AI enclosure. The classification model then processes the gas sensor data and feeds its results to the LLM interface. The AI then initiates a dialogue, sharing its impressions derived from sensor signals, classification outputs, and RAG-retrieved knowledge. These interpretations range from figurative metaphors to relational observations of signal trajectories. Because the AI presents its feedback as a non-authoritative perspective while choices remain editable, participants retain full agency over their judgments, revising or keeping their selection after comparing it with the AI's reading.

Upon confirmation, the system records the judgment and updates both perceivers' diagrams in real time. The mobile application renders the participant's evolving \emph{Genji-mon} pattern, while the AI partner displays its corresponding diagram alongside conversational cues.

\paragraph{Revelation and Debrief.}
In the final phase, the system renders the completed \emph{Genji-mon} patterns on both the mobile interface and the AI partner's display for direct comparison. It then reveals the correct pattern and generates a souvenir bookmark featuring the participant's final diagram. Following the revelation, the AI partner leads a debriefing session explaining how sensor signal trajectories informed its classification results and interpretations.

\section{Pilot Studies}

We conducted two iterative pilot studies.

\paragraph{Pilot Study 1: Framework Validity ($n = 10$).}
The first study evaluated how novice participants without prior olfactory training adapted to the visual notation system of \emph{Genji-mon} patterns. Observations revealed a prevalent anchoring strategy, where participants used the initial scent as a baseline for subsequent comparative judgments. Mapping temporal scent comparisons onto tangible visual patterns provided a stable reference for externalizing abstract sensory data. Overall, the study indicated that the \emph{Genji-k\=o} framework helps non-experts organize and communicate comparative scent impressions.

\paragraph{Pilot Study 2: AI-Guided Reflection ($n = 6$).}
The second study examined how the AI co-smelling partner affected how participants attended to, interpreted, and described the scent sequence. Without presenting its reading as correct, the AI engaged participants through conversational cues, signal-derived descriptions, and figurative metaphors. Participants reported that these alternative perspectives prompted them to revisit and elaborate on their initial sensory impressions. Facing a task that is hard for both parties, participants reported neither over-relying on the AI's interpretations nor dismissing them, instead using the AI's inputs as a helpful reference to re-evaluate the scents and reflect on their choices.

\section{Discussion}

The \emph{Genji-k\=o} framework approaches the olfactory-verbal gap indirectly. Where abstract description in isolation is difficult, it provides a structured anchor that grounds sensory judgments in relative comparison. Articulation thus follows naturally as a means to account for a comparative choice, not as the primary task itself. For interactive systems engaging with subtle sensory modalities, leveraging relational structures offers a viable mechanism to support sensory expression without relying on rigid vocabularies.

The AI partner functions as a co-player whose reading carries no more authority than the participant's. As it shares its signal-derived impressions, the AI also draws on descriptions contributed during prior sessions. This exposes participants to a broader field of ways others have approached the same material, situating a private sensory experience within a shared conversational space. By placing these diverse perspectives alongside human intuition while choices remain editable, the interaction encourages participants to deliberate on their selections. This approach suggests how AI can participate in multi-sensory experiences less as a tool for resolving what is present, and more as a companion that widens the ways sensory impressions are explored and articulated while preserving human agency.

While these insights highlight the potential of AI-mediated sensory dialogue, our approach carries several limitations. Technically, relying on unburnt incense in sealed environments protected sensor integrity, but leaves the system less representative of open-air, combusted incense settings. In addition, the classifier achieves modest accuracy due to overlapping chemical profiles, and the prolonged sampling duration required for data collection introduces frictions that can disrupt the interactional flow. Empirically, our qualitative findings rely on preliminary pilot studies with a small cohort ($n=16$), requiring broader evaluation to confirm long-term perceptual learning and generalizability.

\section{Conclusion and Future Work}

\emph{Smell with Genji} reimagines a traditional scent-matching practice as a medium for human-AI sensory exploration. By coupling the relational structure of \emph{Genji-k\=o} with AI-driven dialogue, our system does not settle what a scent ``is,'' but offers the AI's interpretations while human judgment remains open. That judgment is symbolized by the physical \emph{Genji-mon} pattern the participant carries away. This approach points to how computational systems can foster reflective articulation and preserve human agency in ambiguous sensory domains.

Building on our preliminary observations, future work will proceed along technical and experiential tracks. Technically, we are exploring richer sensor configurations and advanced temporal modeling to better capture subtle nuances among chemically similar materials, enabling more granular AI interpretations. Experientially, we intend to formally evaluate how the linguistic framing of AI feedback, alongside the introduction of diverse personas ranging from analytical to evocative, shapes users' comparative judgments and depth of sensory reflection. 

Additionally, longitudinal studies will help us investigate whether the descriptive strategies and heightened olfactory attention cultivated during these sessions are internalized by users over time. More broadly, the value of olfactory AI may lie less in technical precision than in a co-smelling partnership that opens dialogue about what we perceive.


\end{document}